\documentclass[runningheads]{llncs}
\usepackage[T1]{fontenc}
\usepackage{listings} 
\usepackage{booktabs}
\usepackage{array}
\usepackage{multirow}
\usepackage{url}
\usepackage{graphicx}
\usepackage[table]{xcolor}
\usepackage{amsmath}
\usepackage{tabularx}
\usepackage{makecell}
\usepackage{threeparttable}
\usepackage{hhline}

\begin{document}

\title{Vulnerability Detection in Popular Programming Languages with Language Models
\thanks{This work is partially supported by the Wallenberg AI, Autonomous Systems and Software Program (WASP) funded by the Knut and Alice Wallenberg Foundation and by framework grant RIT17-0032 from the Swedish Foundation for Strategic Research. The computations and data handling were enabled by resources provided by the National Academic Infrastructure for Supercomputing in Sweden (NAISS), partially funded by the Swedish Research Council through grant agreement no. 2022-06725.}}

\author{
    Syafiq Al Atiiq\inst{1} \and
    Christian Gehrmann\inst{1} \and
    Kevin Dahlén\inst{2} 
}
\authorrunning{S. Al Atiiq et al.}
\institute{
    Lund University, Lund, Sweden \\
    \email{\{syafiq\_al.atiiq, christian.gehrmann\}@eit.lth.se} \and
    VyPr AI, Lund, Sweden \\
    \email{kevin@vyprai.net}
}
\maketitle
\begin{abstract}
Vulnerability detection is crucial for maintaining software security, and recent research has explored the use of Language Models (LMs) for this task. While LMs have shown promising results, their performance has been inconsistent across datasets, particularly when generalizing to unseen code. Moreover, most studies have focused on the C/C++ programming language, with limited attention given to other popular languages. This paper addresses this gap by investigating the effectiveness of LMs for vulnerability detection in JavaScript, Java, Python, PHP, and Go, in addition to C/C++ for comparison. We utilize the CVEFixes dataset to create a diverse collection of language-specific vulnerabilities and preprocess the data to ensure quality and integrity. We fine-tune and evaluate state-of-the-art LMs across the selected languages and find that the performance of vulnerability detection varies significantly. JavaScript exhibits the best performance, with considerably better and more practical detection capabilities compared to C/C++. We also examine the relationship between code complexity and detection performance across the six languages and find only a weak correlation between code complexity metrics and the models' F1 scores. 
\end{abstract}
\section{Introduction} 
\label{Intro}
Vulnerability detection is a critical component of software security, as undetected vulnerabilities can lead to severe consequences \cite{1281254}. With the increasing complexity of software, manual detection becomes impractical, requiring the development of automated techniques\footnote{https://lcamtuf.coredump.cx/afl/}. In recent years, deep learning-based approaches have shown promise in vulnerability detection, particularly using language model (LM) \cite{Li_2018}. LMs, such as BERT \cite{devlin2019bertpretrainingdeepbidirectional}, GPT \cite{radford2019language}, and Sonnet \cite{templeton2024scaling}, have demonstrated outstanding performance in code understanding and generation tasks. These models can capture rich semantic and syntactic information from a large codebase, potentially well-suited for analyzing source code \cite{10.1145/3290353}.

Several studies have explored LM's application for vulnerability detection in C/C++ code \cite{9448435,NEURIPS2019_49265d24,10646663}. However, the results have been mixed, with challenges like high false positive rates (FPR) and limited generalization to unseen code \cite{10.1145/3607199.3607242,ding2024vulnerabilitydetectioncodelanguage,atiiq2024generalistspecialistexploringcwespecific}. This raises a question about the effectiveness of LM for in real-world scenarios. While C/C++ is a widely used language, particularly in system-level software, there is a growing need to address the same problem in other programming languages\footnote{\url{https://www.tiobe.com/tiobe-index/}}. Languages such as JavaScript, Java, Python, PHP, and Go are extensively used in web development, enterprise applications, and data analysis\footnote{\url{https://survey.stackoverflow.co/2024/}}. Vulnerabilities in these languages may have significant implications, given their widespread adoption and the sensitive nature of the applications they support\footnote{\url{https://owasp.org/www-project-top-ten/}}.

To bridge this gap, this paper investigates the effectiveness of LMs for vulnerability detection in popular programming languages. Our experimental evaluation covers JavaScript, Java, Python, PHP, and Go. We use the CVEFixes dataset \cite{10.1145/3475960.3475985}, containing a diverse set of vulnerabilities across multiple languages, and preprocess it to create language-specific subsets. These languages are chosen because they have the highest number of samples in the dataset. By fine-tuning and evaluating state-of-the-art (SOTA) language models on these subsets, we assess their performance in detecting vulnerabilities. The paper's main contributions are the following:

\begin{itemize}
    \item We clean and adapt the large CVEFixes dataset to perform an empirical study on the vulnerability detection performance differences between JavaScript, Java, Python, PHP, Go, and C/C++.
    \item We analyze and present correlation figures for the relationship between code complexity and vulnerability detection performance in the investigated dataset.
    \item \sloppy A public release of the curated dataset, scripts, and experimental results to support open science and replicate our findings: \textbf{\url{https://github.com/syafiq/llm_vd}}.
\end{itemize}

The remainder of this paper is organized as follows. Section \ref{sec_bgrw} presents the background and related work on vulnerability detection techniques and language models. Section \ref{problem_definition} provides the problem definition of our work. Section \ref{sec_methodology} describes our dataset, models, and our fine-tuning setup. Section \ref{sec_result} presents the results and analysis of our experiments. Section \ref{threats} discusses the limitations of our work, followed by conclusions and the implications of our findings in section \ref{sec_conclusions}.

\section{Background and Related Work}
\label{sec_bgrw}
\subsection{Vulnerability Detection Techniques}
Over the years, various techniques have been developed to detect vulnerabilities in software systems, ranging from traditional approaches to more advanced deep learning-based methods \cite{9108283,10.1145/3694782,LIANG2025104098}.

\subsubsection{\textbf{Traditional Approaches}}
Traditional vulnerability detection approaches include manual code review \cite{10.1145/3183519.3183525}, static analysis \cite{10.1145/3487019.3487021}, and dynamic analysis \cite{10.1145/318774.318944}. Manual code review is a labor-intensive process where security experts manually inspect the source code to identify vulnerabilities \cite{10.1145/3540250.3549135}. Static analysis techniques examine the source code without executing it to identify potential vulnerabilities \cite{1366126}. Dynamic analysis involves executing the program with specific inputs and monitoring its behavior to detect runtime vulnerabilities. \cite{5504796}.  While these traditional approaches have effectively detected certain vulnerabilities, they have limitations. Manual code review is time-consuming \cite{10.1145/3328778.3367026} and relies heavily on the expertise of the reviewers. Static analysis techniques can suffer from high false positives \cite{tymchuk2017false}, while dynamic analysis may miss vulnerabilities not triggered by the specific inputs used during testing. 

\subsubsection{\textbf{Non-Language Model-based Deep Learning Approaches}}
In recent years, deep learning-based approaches have gained attention for their ability to automatically learn features from large amounts of data and detect complex patterns. These approaches have been applied to various software engineering tasks, including vulnerability detection \cite{9448435,Li_2018,NEURIPS2019_49265d24}. Several studies have explored using deep learning models, such as convolutional neural networks (CNNs) and graph neural networks (GNNs). VulDeePecker \cite{Li_2018} uses CNNs to detect vulnerabilities in C/C++ code. Devign \cite{NEURIPS2019_49265d24} combines GNNs with code property graphs to detect vulnerabilities in C/C++ programs.

While the deep learning-based approaches have shown promising results, they often require extensive feature engineering and may struggle to capture the complex semantics and dependencies present in source code \cite{10.1145/3505243}. Traditional deep learning models typically rely on manual feature engineering to extract relevant information from the code, which can be time-consuming and require domain expertise. Moreover, these models may have difficulty understanding the intricate relationships between code elements, such as variable scoping, control flow, and data dependencies, and capturing the context and intent behind the code. Attention mechanisms used in transformer-based models have been proposed to address these limitations. These mechanisms can help capture long-range dependencies and global context. However, even with these advancements, deep learning models may still struggle to fully capture source code's complex semantics and nuances, especially when dealing with diverse programming languages and coding styles. 

\subsection{Vulnerability Detection with Language Model}
The language model (LM) is a class of deep learning models trained on vast amounts of text data to learn rich natural language representations. These models, such as BERT \cite{devlin2019bertpretrainingdeepbidirectional}, GPT \cite{radford2019language}, Gemini \cite{geminiteam2024geminifamilyhighlycapable}, and Deepseek \cite{dai2024deepseekmoeultimateexpertspecialization}, have achieved state-of-the-art performance on a wide range of natural language tasks. LMs are typically based on transformers \cite{NIPS2017_3f5ee243}, which use self-attention mechanisms to capture long-range dependencies in the input. By pre-training on large corpora of text data, LM learns to understand the structure and semantics of natural language, enabling them to generate coherent and contextually relevant text. The success of LM in natural language has sparked interest in applying them to other domains, such as source code analysis and generation. Several studies have explored the use of LM for tasks such as code completion \cite{svyatkovskiy2020intellicodecomposecodegeneration}, code summarization \cite{feng-etal-2020-codebert}, and bug detection \cite{10.1145/3649828}.

\subsubsection{\textbf{LM for Code Analysis}}
Source code, like natural language, shows rich structural and semantic properties. LM, with their ability to capture complex patterns, have shown promise in analyzing source code \cite{10.1145/3290353,10.1145/3597503.3639187,shi2024naturallanguageoutlinescode}. Several LMs have been specifically designed for code analysis. CodeBERT \cite{feng-etal-2020-codebert} is a pre-trained model that learns from a large corpus of programming language data, enabling it to understand the semantics of code. GraphCodeBERT \cite{guo2021graphcodebertpretrainingcoderepresentations} extends CodeBERT by incorporating graph-based representations of code, capturing the structural information present in abstract syntax trees and data flow graphs.

\subsubsection{\textbf{Vulnerability Detection using LM}}

The application of LM for vulnerability detection has gained attention in recent years, mainly in the context of C/C++ code. Several studies investigated the effectiveness of LM in detecting vulnerabilities and reported promising results and significant challenges. 

In C/C++, two notable datasets have been introduced. DiverseVul \cite{10.1145/3607199.3607242} is a dataset containing 18,945 vulnerable functions spanning 150 CWEs and 330,492 non-vulnerable functions. The authors study 11 model architectures from 4 families (GNN \cite{4700287}, RoBERTa \cite{liu2019roberta}, GPT-2 \cite{radford2019language}, and T5 \cite{wang2021codet5}) and find that LM outperforms GNN, especially when trained on larger datasets. However, the model still struggles with high FPR and limited generalization. As a follow-up, PrimeVul \cite{ding2024vulnerabilitydetectioncodelanguage} is another C/C++ dataset designed to address the limitations of existing datasets, i.e., poor data quality, low label accuracy, and high duplication rates. The authors evaluate various LMs on PrimeVul and find that their performances are significantly lower than reported for the previous datasets. The performance remains poor even with advanced training techniques and larger models.

The major issue with the detection results reported from the PrimeVul \cite{ding2024vulnerabilitydetectioncodelanguage} and DiverseVul \cite{10.1145/3607199.3607242} is is the low F1 scores. PrimeVul highlights the significant underperformance of LM when faced with realistic, diverse, and challenging vulnerabilities, even with advanced training techniques and SOTA models. Similarly, the DiverseVul study pinpoints a significant challenge for models to generalize to unknown test projects on the vulnerability detection task.

Furthermore, the vulnerability types diversity, each with its own unique characteristics, code semantics, and patterns, makes it challenging for a single binary classifier to learn and detect vulnerabilities across all categories effectively \cite{atiiq2024generalistspecialistexploringcwespecific}. The imbalance between vulnerable and non-vulnerable code samples in datasets can also lead to biased models that struggle to learn and detect vulnerabilities effectively \cite{10.1145/3607199.3607242,ding2024vulnerabilitydetectioncodelanguage}. Inadequate training techniques have also been identified as a limitation. Advanced training techniques, such as class weights and contrastive learning (both performed in \cite{10.1145/3607199.3607242,ding2024vulnerabilitydetectioncodelanguage}), have shown limited success in improving the performance. Fundamentally, new approaches may be needed to address this.

PrimeVul highlights key areas where the current LM falls short. First, prior work focuses on function-level analysis without considering the broader context, such as interprocedural data flows, making detecting vulnerabilities challenging even for humans. Second, LM makes decisions primarily based on textual similarity without considering the underlying root causes. Finally, posing vulnerability detection as a binary classification problem might be oversimplistic. A more nuanced approach that decomposes the problem into sub-problems and teaches the model to reason about each step might be more effective.

\section{Problem Definition}
\label{problem_definition}

As shown in our review of previous results in Section \ref{sec_bgrw}, while LM looks promising in vulnerability detection for C/C++ code, significant challenges remain, including low F1 scores, limited generalization, vulnerability types diversity, data imbalance, and inadequate training techniques. The models built for C/C++ did not work as expected, spurring the need to explore whether similar problems appear for other programming languages. Despite this fact, few efforts have been devoted to investigating if similar issues exist for other programming languages or even if there are large detection performance differences between languages. We address this research gap by making an experimental evaluation of the differences in vulnerability detection using LMs for JavaScript, Java, Python, PHP, and Go. We also include figures with C/C++ to make a proper comparison between our results and the previous results on C/C++ possible. Proper comparison can only be done if a reasonably similar dataset for all languages is available. To address this issue, we present a cleaned new dataset that can be used to make sound comparisons. With the help of such a dataset, the related research question also arises: ``If we have performance differences between different languages, can this difference be due to code complexity variations?''. We address this research question as well, presenting figures on the correlation between vulnerability detection performance and code complexity for the given dataset. 

\section{Dataset, Models and Setup}
\label{sec_methodology}
\subsection{Dataset}
Several datasets have been proposed for building vulnerability detection systems, including: (i) C/C++ based datasets such as DiverseVul \cite{10.1145/3607199.3607242} and PrimeVul \cite{ding2024vulnerabilitydetectioncodelanguage}; (ii) project-specific datasets like ReVeal \cite{9448435}, which focuses on the Linux Debian kernel and Chromium projects; and (iii) cross-language datasets such as CrossVul \cite{10.1145/3468264.3473122} and CVEFixes \cite{10.1145/3475960.3475985}. These datasets provide diverse vulnerabilities and have been widely used in the research community. Among these, only CrossVul and CVEFixes are suitable for our purpose. Both CrossVul and CVEFixes extract the data based on the same Common Vulnerabilities and Exposures (CVE) records in the public U.S. National Vulnerability Database (NVD)\footnote{\url{https://nvd.nist.gov/}}, so using both would create unnecessary duplication. However, only CVEFixes provides the data update regularly (as well as an open-source script\footnote{\url{https://github.com/secureIT-project/CVEfixes}} to scrape the data on our own if the newest data is unavailable yet). After considering the requirements of our study, including the need for cross-language vulnerability data and regular updates, we ultimately selected CVEFixes as the most suitable dataset to serve as the foundation of our work.

\begin{figure}[ht]
\centering
\includegraphics[width=0.7\columnwidth]{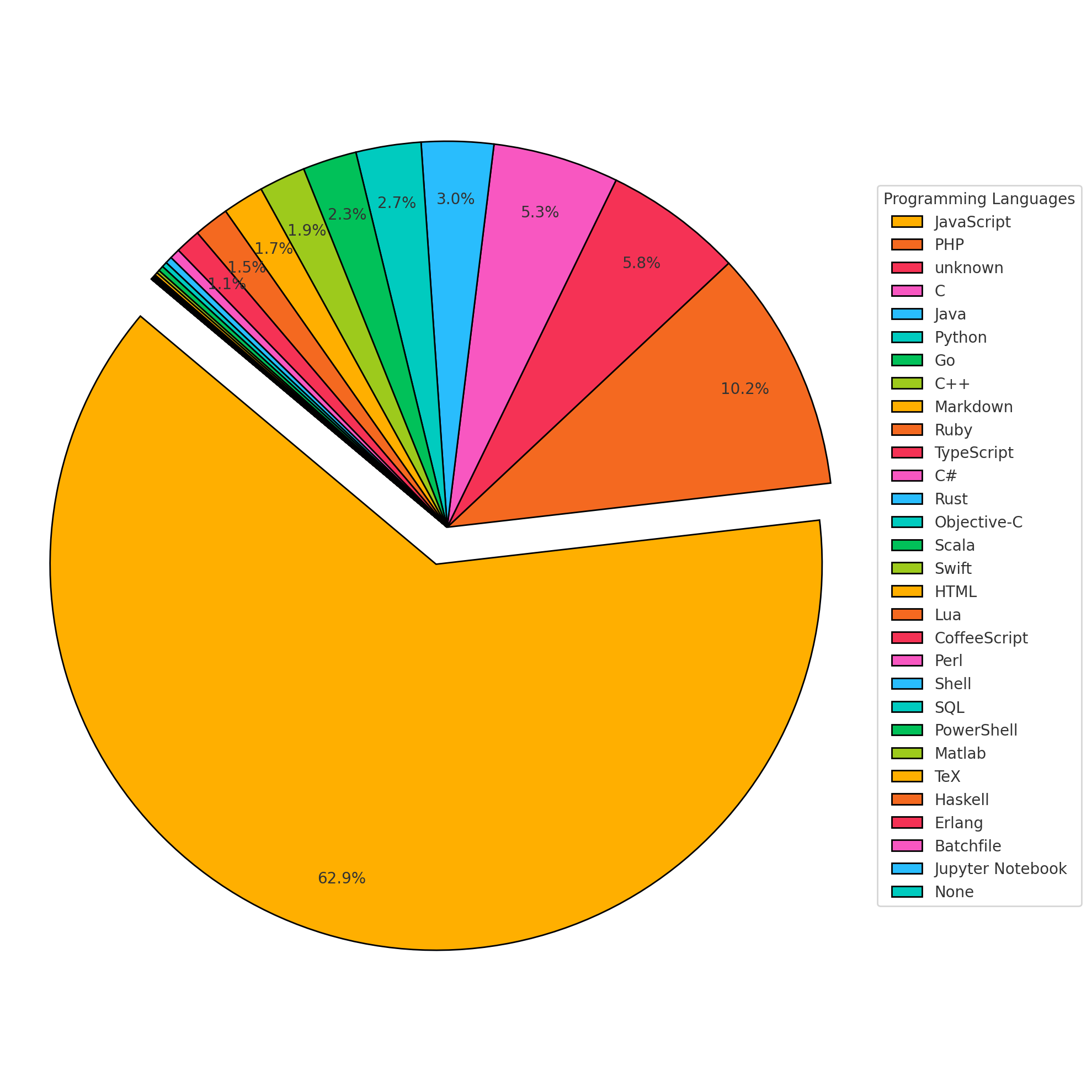}
\caption{Distribution of programming languages in the CVEfixes dataset.}
\label{fig:language_distribution}
\end{figure}

CVEfixes initial release spans all published CVE records up to June 9, 2021, covering 5365 CVE records for 1754 open-source projects. 5495 vulnerability fixing commits are obtained from the projects' version control systems and linked to information from the corresponding CVE records, such as CVE-IDs, reference links, severity scores, CWE type, and other descriptive information. The latest release covers all published CVEs up to 23 July 2024. Figure \ref{fig:language_distribution} shows the distribution of programming languages in the CVEFixes dataset based on the latest scraped data from 2024. The dataset consists of 277,948 entries, with 126,599 classified as vulnerable and 151,349 as non-vulnerable. It covers many languages, with JavaScript being the most prevalent, followed by PHP, Java, Python, and Go. These top 5 languages (excluding C, C++, and unknown) were selected as the focus of our study due to their significant representation.

To validate the generalizability of our findings, we tested our models on independent datasets where available. For Java, we used the MegaVul dataset \cite{10555623} (only the Java part\footnote{\url{https://github.com/Icyrockton/MegaVul}}). For Python, we tested on the synth-vuln-fixes dataset\footnote{\url{https://huggingface.co/datasets/patched-codes/synth-vuln-fixes}} - a curated dataset of Python vulnerabilities generated by GPT-4 and validated through both human review and static analysis. For PHP, we used the PHP vulnerability test suite from SARD\footnote{\url{https://samate.nist.gov/SARD/test-suites/103}} (Software Assurance Reference Dataset) \cite{7515499}. It's important to note that the SARD PHP test cases are synthetic, meaning they were specifically created as examples with well-characterized weaknesses. We were unable to find suitable alternative datasets containing vulnerabilities and their fixes for JavaScript and Go. For languages where we had alternative datasets, this cross-dataset validation helps assess how well our models generalize beyond the CVEFixes data.

\subsubsection{\textbf{Data Preprocessing and Filtering}}
We first split CVEfixes based on the programming language. For each subset, we perform the following preprocessing steps:
\begin{enumerate}
    \item Duplication removal: We identify and remove any duplicate entries in the dataset to avoid data leakage \cite{rosenblatt2024data}.
    \item Train-test splitting: To maintain temporal integrity and simulate a realistic evaluation scenario, we divide the data into training and testing sets based on the commit timestamps using stratified sampling. Stratified sampling is a technique that ensures each class (in this case, vulnerable and non-vulnerable) is proportionally represented in both the training and testing sets. This is particularly important when dealing with imbalanced datasets, as it helps maintain the original class distribution and prevents bias towards the majority class. After stratifying the data, we split it so that the model is trained on data from earlier timestamps and evaluated on data from later timestamps. This method ensures that the model's performance is assessed on future, unseen vulnerabilities, thereby avoiding any inadvertent leakage of future information into the training process. This also tests the model's ability to catch vulnerabilities that have not been seen before.
\end{enumerate}

Table \ref{table:vulnerability_counts} shows the distribution of the five subsets (and C/C++ for comparison) used in this paper.

\begin{table}[h!]
\centering
\scriptsize
\begin{tabular}{l|r|r|r}
\hline
\textbf{Language} & \textbf{Vulnerable} & \textbf{Non-vulnerable} & \textbf{Total} \\
\hline
JavaScript & 46,802 (82,999) & 53,198 (91,929) & 100,000 (174,928) \\
\hline
PHP         & 4,758  & 23,499  & 28,257 \\
\hline
C/C++       & 8,299  & 11,761  & 20,060 \\
\hline
Java        & 3,102  & 5,285   & 8,387  \\
\hline
Python      & 2,775  & 4,793   & 7,568  \\
\hline
Go          & 1,880  & 4,403   & 6,283  \\
\hline
\end{tabular}
\caption{Vulnerable and non-vulnerable counts by programming language. Note: for JavaScript, we only used the first 100,000 entries (57.17\% of the total data).}
\label{table:vulnerability_counts}
\end{table}

\subsection{Models}

\begin{table}
    \centering
    \scriptsize
    \begin{tabular}{c|c|c|c|c}
        \hline
        \textbf{Model} & \textbf{Acronym} & \textbf{Parameters} & \textbf{Arch} & \textbf{Method} \\
        \hline
        CodeT5 \cite{wang2021codet5} & CT5 & 60 M & Encoder-Decoder & Fine-tune \\
        \hline
        CodeBERT \cite{feng-etal-2020-codebert} & CB & 125 M & Encoder & Fine-tune \\
        \hline
        UnixCoder \cite{guo-etal-2022-unixcoder} & UC & 125 M & Encoder & Fine-tune \\
        \hline
        DeepSeek-Coder \cite{deepseek-coder} & DS-C & 1.3 B & Decoder & Fine-tune \\
        \hline
    \end{tabular}
    \caption{LMs we study.}
    \label{tab_listlm}
\end{table}

We investigate the performance of the language-specific subset data for the vulnerability detection task with four LMs. Table \ref{tab_listlm} presents the four LMs evaluated in this study, along with their parameter sizes and architectures. DeepSeek-Coder, with 1.3 billion parameters, is the largest model among the four, followed by CodeBERT and UnixCoder, both having 125 million parameters, and finally, CodeT5, with 60 million parameters. We perform full fine-tuning on all the models and evaluate them independently.

\subsection{Experimental Setup}

We design the experiments with all the LMs following the existing benchmarks \cite{10.1145/3607199.3607242,ding2024vulnerabilitydetectioncodelanguage,lu2021codexglue}. This includes setting the learning rate to be $2\times10^{-5}$, running the fine-tuning for ten epochs, and a batch size of one per GPU. We employ the AdamW optimizer and use gradient accumulation with a step size of 8 to simulate a larger effective batch size. The learning rate is scheduled using a linear warmup for 1000 steps followed by a linear decay. The LMs are fine-tuned using the PyTorch deep learning framework on a node of NVIDIA A100 GPUs.

\section{Results and Analysis}
\label{sec_result}

We aim to evaluate the performance of LMs in detecting vulnerabilities across various popular programming languages. This is particularly intriguing given that recent research findings on C/C++ indicate significant challenges related to detection performance in real-world applications, as discussed in Section \ref{Intro}.

\subsection{Evaluation Metrics}
We employ several evaluation metrics to assess the performance:
\begin{itemize}
\item \textbf{Accuracy}: The proportion of correctly classified samples (both vulnerable and non-vulnerable) out of the total number of samples.
\item \textbf{Precision}: The proportion of correctly identified vulnerable samples out of all samples classified as vulnerable by the model.
\item \textbf{Recall}: The proportion of correctly identified vulnerable samples out of all actual vulnerable samples in the dataset.
\item \textbf{F1 Score}: The harmonic mean of precision and recall, providing a balanced measure of the model's performance.
\item \textbf{False Positive Rate (FPR)}: The proportion of non-vulnerable samples incorrectly classified as vulnerable out of all non-vulnerable samples.
\end{itemize}

\subsection{Language-dependent Detection Performance.}
\label{result_RQ1}

\begin{table*}[h!]
    \centering
    \scriptsize
    \begin{threeparttable}
        \begin{tabular}{|c|c|c|c|c|c|c|c|c|}
             \hline
             Language & Model & Train & Test & Acc $\uparrow$ & F1 $\uparrow$ & Prec $\uparrow$ & Rec $\uparrow$ & FPR $\downarrow$ \\
             \hline
             \multirow{4}{*}{JavaScript$^{1}$} 
             & CT5 & \multirow{4}{*}{CVEFixes} & \multirow{4}{*}{CVEFixes} & 50.53 & 27.17 & 50.80 & 18.54 & 17.79 \\
             \cline{2-2}\cline{5-9}
             & CB & & & 71.46 & 69.25 & 74.62 & 64.60 & 21.75 \\
             \cline{2-2}\cline{5-9}
             & UC & & & 70.11 & 70.23 & 69.61 & 70.86 & 30.64 \\
             \cline{2-2}\cline{5-9}
             & DS-C & & & 69.59 & 65.55 & 75.13 & 58.14 & 19.07 \\
             \hline
             \multirow{8}{*}{PHP$^{1}$} 
             & CT5 & \multirow{8}{*}{CVEFixes} & \multirow{4}{*}{CVEFixes} & 82.30 & 24.32 & 27.10 & 22.05 & 8.78 \\
             \cline{2-2}\cline{5-9}
             & CB & & & 81.62 & 28.84 & 28.79 & 28.90 & 10.58 \\
             \cline{2-2}\cline{5-9}
             & UC & & & 81.67 & 38.08 & 33.72 & 43.73 & 12.72 \\
             \cline{2-2}\cline{5-9}
             & DS-C & & & 82.84 & 36.59 & 34.95 & 38.40 & 10.58 \\
             \cline{2-2}\cline{4-9}
             & CT5 & & \multirow{4}{*}{SARD} & 47.80 & 46.41 & 33.78 & 74.10 & 63.74 \\
             \cline{2-2}\cline{5-9}
             & CB & & & 38.50 & 39.41 & 28.17 & 65.57 & 73.38 \\
             \cline{2-2}\cline{5-9}
             & UC & & & 53.10 & 17.86 & 19.17 & 16.72 & 30.93 \\
             \cline{2-2}\cline{5-9}
             & DS-C & & & 37.80 & 25.42 & 20.04 & 34.75 & 60.86 \\
             \hline
             \multirow{8}{*}{Java$^{1}$} 
             & CT5 & \multirow{8}{*}{CVEFixes} & \multirow{4}{*}{CVEFixes} & 53.98 & 62.75 & 52.30 & 78.41 & 69.90 \\
             \cline{2-2}\cline{5-9}
             & CB & & & 57.22 & 65.54 & 54.45 & 82.30 & 67.30 \\
             \cline{2-2}\cline{5-9}
             & UC & & & 57.83 & 65.47 & 54.99 & 80.88 & 64.71 \\
             \cline{2-2}\cline{5-9}
             & DS-C & & & 68.59 & 69.18 & 67.17 & 71.33 & 34.08 \\
             \cline{2-2}\cline{4-9}
             & CT5 & & \multirow{4}{*}{MegaVul} & 59.75 & 66.56 & 56.93 & 80.10 & 60.60 \\
             \cline{2-2}\cline{5-9}
             & CB & & & 58.80 & 64.27 & 56.74 & 74.10 & 56.50 \\
             \cline{2-2}\cline{5-9}
             & UC & & & 65.65 & 70.11 & 62.04 & 80.60 & 49.30 \\
             \cline{2-2}\cline{5-9}
             & DS-C & & & 57.20 & 63.88 & 55.26 & 75.70 & 61.30 \\
             \hline
             \multirow{8}{*}{Python$^{1}$} 
             & CT5 & \multirow{8}{*}{CVEFixes} & \multirow{4}{*}{CVEFixes} & 54.89 & 45.57 & 41.87 & 50.00 & 42.14 \\
             \cline{2-2}\cline{5-9}
             & CB & & & 59.24 & 47.74 & 46.28 & 49.28 & 34.72 \\
             \cline{2-2}\cline{5-9}
             & UC & & & 55.03 & 52.24 & 43.61 & 65.11 & 51.09 \\
             \cline{2-2}\cline{5-9}
             & DS-C & & & 60.19 & 53.27 & 47.85 & 60.07 & 39.74 \\
             \cline{2-2}\cline{4-9}
             & CT5 & & \multirow{4}{*}{synth-vuln-fixes} & 55.47 & 55.58 & 55.45 & 55.72 & 44.78 \\
             \cline{2-2}\cline{5-9}
             & CB & & & 22.14 & 0.63 & 0.87 & 0.49 & 56.22 \\
             \cline{2-2}\cline{5-9}
             & UC & & & 35.82 & 40.55 & 37.77 & 43.78 & 72.13 \\
             \cline{2-2}\cline{5-9}
             & DS-C & & & 41.04 & 1.66 & 5.00 & 0.99 & 18.91 \\
             \hline
             \multirow{4}{*}{Go$^{1}$} 
             & CT5 & \multirow{4}{*}{CVEFixes} & \multirow{4}{*}{CVEFixes} & 89.19 & 37.25 & 29.23 & 51.35 & 8.29 \\
             \cline{2-2}\cline{5-9}
             & CB & & & 86.66 & 43.17 & 29.41 & 81.08 & 12.97 \\
             \cline{2-2}\cline{5-9}
             & UC & & & 92.57 & 55.10 & 44.26 & 72.97 & 6.12 \\
             \cline{2-2}\cline{5-9}
             & DS-C & & & 85.30 & 41.61 & 27.68 & 83.78 & 14.59 \\
             \hline
             \multirow{8}{*}{C/C++$^{1}$} 
             & CT5 & \multirow{8}{*}{CVEFixes} & \multirow{4}{*}{CVEFixes} & 57.13 & 33.06 & 45.25 & 26.05 & 21.58 \\
             \cline{2-2}\cline{5-9}
             & CB & & & 57.59 & 42.71 & 47.37 & 38.88 & 29.59 \\
             \cline{2-2}\cline{5-9}
             & UC & & & 59.50 & 1.25 & 71.43 & 0.63 & 0.17 \\
             \cline{2-2}\cline{5-9}
             & DS-C & & & 60.02 & 34.96 & 51.61 & 26.43 & 16.97 \\
             \cline{2-2}\cline{4-9}
             & CT5 & & \multirow{4}{*}{PrimeVul} & 71.28 & 3.86 & 2.09 & 26.05 & 27.69 \\
             \cline{2-2}\cline{5-9}
             & CB & & & 48.51 & 3.11 & 1.62 & 37.34 & 51.23 \\
             \cline{2-2}\cline{5-9}
             & UC & & & 96.97 & 1.57 & 2.79 & 1.09 & 0.86\\
             \cline{2-2}\cline{5-9}
             & DS-C & & & 65.55 & 2.02 & 1.08 & 16.03 & 33.33 \\
             \hhline{|=|=|=|=|=|=|=|=|=|}
             \multirow{2}{*}{C/C++$^{2}$} 
             & CT5 & \multirow{2}{*}{DiverseVul} & \multirow{2}{*}{DiverseVul} & 94.91 & 9.39 & 13.35 & 7.24 & 1.78 \\
             \cline{2-2}\cline{5-9}
             & CB & & & 94.19 & 11.94 & 13.34 & 10.80 & 2.65 \\
             \hline
             \multirow{3}{*}{C/C++$^{2}$} 
             & CT5 & \multirow{3}{*}{PrimeVul} & \multirow{3}{*}{PrimeVul} & 96.67 & 19.7 & - & - & - \\
             \cline{2-2}\cline{5-9}
             & CB & & & 96.87 & 20.86 & - & - & - \\
             \cline{2-2}\cline{5-9}
             & UC & & & 96.86 & 21.43 & - & - & - \\
             \hline
             \multirow{2}{*}{C/C++$^{2}$} 
             & \multirow{2}{*}{DS-C} & \multirow{2}{*}{Gen2Spec} & Gen2Spec & 75.68 & 23.36 & 13.76 & 77.35 & 24.41  \\
             \cline{4-9}
             & & & PrimeVul & 76.46 & 14.58 & 7.93 & 90.71 & 23.86  \\
             \hline
        \end{tabular}
        \caption{The performance of LMs (using acronyms from table \ref{tab_listlm}) to the respective programming languages.}
        \label{tab_allresult}
        \begin{tablenotes}
            \item[1] Measured by us.
            \item[2] Taken from the respective paper with the size of the datasets can be seen in table \ref{tab_prev_dataset}. (-) means no data.
        \end{tablenotes}
    \end{threeparttable}
\end{table*}

\begin{table}
    \centering
    \scriptsize
    \begin{tabular}{|c|c|c|c|}
         \hline
         Dataset & Vulnerable & Non-Vulnerable & Total \\
         \hline         
         DiverseVul \cite{10.1145/3607199.3607242} & 18,945 & 311,997 & 330,492 \\
         \hline
         PrimeVul \cite{ding2024vulnerabilitydetectioncodelanguage} & 6,968 & 228,800 & 235,768 \\
         \hline
         Gen2Spec \cite{atiiq2024generalistspecialistexploringcwespecific} & 16,955 & 295,587 & 312,542 \\
         \hline
         C/C++@CVEFixes & 8,299  & 11,761  & 20,060 \\
         \hline
    \end{tabular}
    \caption{Size of previous C/C++ datasets.}
    \label{tab_prev_dataset}
\end{table}

The analysis of vulnerability detection performance from Table \ref{tab_allresult} in models trained on non-C/C++ languages (JavaScript, PHP, Java, Python, and Go) reveals varying levels of effectiveness. These models achieve accuracy ranging from 50\% to 93\% and F1 scores spanning from 24\% to 70\%, indicating moderate to good performance in identifying vulnerabilities in these languages. However, the FPRs for non-C/C++ languages are often higher, ranging from 6\% to 69\%, depending on the language and dataset used. The higher FPR suggests that while these models are generally capable of detecting vulnerabilities, they also tend to incorrectly flag more code snippets as vulnerable, which could lead to unnecessary manual review efforts.

The higher FPR scores observed can be attributed to several factors. While the CVEfixes dataset used in our study covers a significant portion (80\%) of the available vulnerability fixes from the NVD database, the dataset sizes for individual languages like Go, Java, and Python are relatively smaller compared to the C/C++ datasets used in previous studies. However, it is important to note that these smaller dataset sizes are a result of the inherent distribution of vulnerabilities across languages in the NVD database (reported real-world representation) and not a limitation of the CVEfixes dataset itself.

In addition to dataset size, other factors such as dataset quality, diversity, and the inherent characteristics of the programming languages may also contribute to the observed differences in performance. The relatively smaller datasets for languages like Go, Java, and Python in the current study may impact the models' ability to learn and generalize effectively, leading to higher FPR scores despite their overall better performance in terms of F1 score compared to the results obtained in DiversVul and PrimeVul for C/C++. Even if this is the case, the datasets used in our comparison are rather similar, allowing a better comparison between the models under the same conditions. This indicates a better ability to detect vulnerabilities with LLM for JavaScript compared to C/C++.

\sloppy Among the non-C/C++ languages, the performance varies. JavaScript, Java, and Python perform better overall than PHP and Go. Models trained on JavaScript, Java, and Python achieve moderate accuracy, ranging from 50\% to 71\%, and F1 scores between 27\% and 70\%. These results suggest that the models are relatively effective in identifying vulnerabilities in these languages, although there is still room for improvement.

For PHP, the high accuracy (81\% to 83\%) and low F1 scores (24\% to 38\%) could result from the significant class imbalance in the dataset, with around 4,758 vulnerable samples and 23,499 non-vulnerable samples. Despite the stratified train-test splitting approach, the limited number of vulnerable samples in the training set might hinder the models' ability to learn effective patterns for detecting vulnerabilities, leading to a bias toward classifying most samples as non-vulnerable. In the case of Go, the models show high accuracy (85\% to 93\%) but low F1 scores (37\% to 55\%). While Go has a more balanced dataset compared to PHP, the models still struggle to detect vulnerabilities effectively. This suggests that vulnerability patterns in Go might be more complex or harder to learn with the current model architectures.

The performance of different LMs varies across programming languages, suggesting that no single model architecture consistently outperforms others in all scenarios. Interestingly, the performance rankings of these models do not strictly align with their parameter sizes, showing that factors other than model size contribute to their effectiveness in vulnerability detection.

UnixCoder and CodeBERT, despite having fewer parameters than DeepSeek-Coder, consistently demonstrate strong performance across multiple languages. On the other hand, CodeT5, the smallest model among the four, often lags behind other models in terms of performance across different languages. This indicates that while a larger model size can potentially improve performance, it is not the sole determining factor. DeepSeek-Coder, the largest model, shows mixed performance results across languages. While it achieves the highest accuracy and F1 scores for some languages, such as Java, it does not consistently outperform other models in all cases. This suggests that the relationship between model size and performance is not always linear.

\subsubsection{\textbf{Validation on Other Dataset}}

To validate the generalizability of the models, it is essential to test their performance on independent datasets. In this study, we evaluated the models trained on the CVEFixes dataset on several independent datasets for different languages.

For Java, the models were tested on the MegaVul dataset. The results show slightly lower performance compared to their performance on the CVEFixes dataset. However, the performance remains consistent across both datasets, with the models maintaining their relative rankings. This consistency in performance provides some validation of the results and suggests that the models are capable of generalizing to new Java vulnerability data to a certain extent.

For PHP, we tested the models on the PHP vulnerability test suite from SARD. The results indicate a drop in performance when compared to the models' performance on the CVEFixes dataset, particularly in terms of Precision and Recall. Additionally, the FPR increased significantly. This can be attributed to the fact that the SARD PHP test cases are synthetic, meaning they were specifically created as examples with well-characterized weaknesses. Each test case targets only one specific flaw. As a result, these test cases have a much simpler structure than most weaknesses found in real-world production code. The models trained on the CVEFixes PHP subset likely learned patterns and characteristics specific to the real-world vulnerabilities present in that dataset, making it challenging for them to generalize to the synthetic, targeted vulnerabilities in the SARD test cases.

In the case of Python, we evaluated the models on the synth-vuln-fixes dataset. The results show mixed performance, with some models (i.e., CodeT5) maintaining relatively consistent performance across both datasets, while others (i.e., CodeBERT and DeepSeek-Coder) exhibit a significant drop in performance. This indicates that the ability of the models to generalize to new Python vulnerability data may depend on the specific model architecture.

On the other hand, the poor performance of the models in the C/C++ language persists across multiple datasets, including CVEFixes, PrimeVul, DiverseVul, and Gen2Spec. Despite the differences in the datasets' characteristics and the vulnerability types they cover, the models consistently struggle to effectively detect vulnerabilities in C/C++ code. For C/C++, the performance when testing on CVEFixes is better than testing on PrimeVul and other C/C++ datasets, as those datasets have more extensive entries compared to the C/C++ subset in CVEFixes, as shown in Table \ref{tab_prev_dataset}. This suggests that the larger dataset sizes in PrimeVul, DiverseVul, and Gen2Spec may contribute to the lower performance observed when testing on these datasets compared to CVEFixes. However, the overall poor performance across all datasets highlights the inherent difficulty of vulnerability detection in C/C++ and suggests that the challenges are not limited to a specific dataset but are rather a fundamental problem in applying current language models to C/C++ vulnerability detection.

It would be natural to assume that the lower performance of vulnerability detection models for C/C++ compared to other languages may be attributed to the inherent complexity of C/C++ code, as these languages are commonly used in system-level software development. The low-level nature of C/C++, manual memory management, and the potential for more diverse coding styles may introduce challenges for the models to generalize across different codebases. These facts motivate our following second analysis of code complexity versus detection performance.

\subsection{Code Complexity and Detection Performance}
\label{result_RQ2}

In this section, we investigate the relationship between code complexity and vulnerability detection performance across the languages we assess. The detection performance is measured using the F1 score of all the LMs for each language. Specifically, we aim to determine whether higher code complexity correlates with lower detection performance, as measured by the model's F1 score. The choice of F1 score is motivated by several factors. First, the F1 score has been widely used in previous vulnerability detection studies, including those focusing on C/C++ \cite{10.1145/3607199.3607242,9448435,ding2024vulnerabilitydetectioncodelanguage}. By using the F1 score, we maintain consistency with prior work and facilitate comparisons across different studies. Second, vulnerability detection datasets often have imbalanced classes (this one applies to our dataset as well, even though the imbalance is not that high, see table \ref{table:vulnerability_counts}), with a higher proportion of non-vulnerable samples compared to vulnerable ones. F1 score is less sensitive to class imbalance than other metrics like accuracy, making it a more suitable choice for evaluating vulnerability detection systems.

To assess code complexity, we compute several established metrics:

\begin{itemize} 
\item \textbf{Token Length}: The average number of tokens per code snippet. 
\item \textbf{Halstead Volume}: Measures the size of the implementation of an algorithm. 
\item \textbf{Halstead Difficulty}: Indicates the difficulty level of writing or understanding the code. 
\item \textbf{Halstead Effort}: Estimates the effort required to implement or understand the code. 
\item \textbf{Cyclomatic Complexity}: Quantifies the number of linearly independent paths through the code, reflecting its control flow complexity. 
\item \textbf{Number of Lines of Code (NLOC)}: The average number of lines per code snippet. 
\end{itemize}

Table~\ref{tab_complexity_metrics} presents the mean values of these complexity metrics for each programming language we analyze.

\begin{table}
    \centering
    \scriptsize
    \begin{tabular}{@{}lcccccc@{}}
        \toprule
        \textbf{Language} & 
        \makecell{\textbf{Token} \\ \textbf{Length}} & 
        \makecell{\textbf{Halstead} \\ \textbf{Volume}} & 
        \makecell{\textbf{Halstead} \\ \textbf{Difficulty}} & 
        \makecell{\textbf{Halstead} \\ \textbf{Effort}} & 
        \makecell{\textbf{Cyclomatic} \\ \textbf{Complexity}} & 
        \textbf{NLOC} \\
        \midrule
        JavaScript  & 462.05 & 2204.25 & 21.83 & 76395.58 & 9.46 & 6.39 \\
        PHP         & 244.44 & 5.79    & 0.62  & 105.03    & 2.02 & 1.22 \\
        Java        & 274.79 & 879.30  & 10.81 & 16462.39  & 3.27 & 20.50 \\
        Python      & 338.35 & 862.93  & 13.07 & 20805.42  & 4.36 & 23.43 \\
        Go          & 752.59 & 2118.29 & 23.86 & 78381.61  & 7.89 & 53.57 \\
        C/C++       & 274.79 & 879.30  & 10.81 & 16462.39  & 3.27 & 20.50 \\
        \bottomrule
    \end{tabular}
    \caption{Mean code complexity metrics by programming language.}
    \label{tab_complexity_metrics}
\end{table}

\subsubsection{\textbf{Correlation Analysis}}
We compute the Pearson correlation coefficients between each code complexity metric and the F1 score for each model to assess the strength and direction of their relationships. Table~\ref{tab_correlation_coefficients} summarizes the correlation coefficients and their corresponding $p$-values.

The correlation coefficients vary in both magnitude and sign across the different models and complexity metrics. For the CodeT5 model, most of the coefficients are negative, indicating a slight inverse relationship between code complexity and F1 score, with the exception of NLOC, which shows a positive correlation. However, for the other models (CodeBert, UnixCoder, and  DeepSeek-Coder), the coefficients are mostly positive, suggesting a slight positive relationship between code complexity and F1 score. Notably, though, the absolute values of all the coefficients are relatively low, generally below 0.6. Furthermore, the $p$-values for all correlations are much higher than the typical significance level of 0.05, indicating that none of these correlations are statistically significant. This suggests that the observed relationships between the various code complexity measures and model performance could be due to random chance rather than a true underlying correlation.

\begin{table}[ht]
    \centering
    \scriptsize
    \renewcommand{\arraystretch}{1.2}
    \label{tab:correlation_metrics}
    \begin{tabular}{|l|l|c|c|}
        \hline
        \textbf{Model} & \textbf{Complexity Metric} & \textbf{Correlation Coefficient (r)} & \textbf{P-value} \\
        \hline
        \multirow{6}{*}{\textbf{CT5}} 
        & Token Length          & -0.1271 & 0.8104 \\
        & Halstead Volume       & -0.0676 & 0.8988 \\
        & Halstead Difficulty   &  0.0363 & 0.9456 \\
        & Halstead Effort       & -0.2002 & 0.7037 \\
        & Cyclomatic Complexity & -0.2347 & 0.6544 \\
        & NLOC                  &  0.3197 & 0.5368 \\
        \hline
        \multirow{6}{*}{\textbf{CB}} 
        & Token Length          &  0.0786 & 0.8823 \\
        & Halstead Volume       &  0.5624 & 0.2453 \\
        & Halstead Difficulty   &  0.5288 & 0.2808 \\
        & Halstead Effort       &  0.4339 & 0.3900 \\
        & Cyclomatic Complexity &  0.5177 & 0.2928 \\
        & NLOC                  & -0.0681 & 0.8980 \\
        \hline
        \multirow{6}{*}{\textbf{UC}} 
        & Token Length          &  0.3619 & 0.4808 \\
        & Halstead Volume       &  0.4558 & 0.3637 \\
        & Halstead Difficulty   &  0.4371 & 0.3861 \\
        & Halstead Effort       &  0.4853 & 0.3292 \\
        & Cyclomatic Complexity &  0.5336 & 0.2756 \\
        & NLOC                  &  0.0555 & 0.9169 \\
        \hline
        \multirow{6}{*}{\textbf{DS-C}} 
        & Token Length          & -0.0431 & 0.9354 \\
        & Halstead Volume       &  0.3435 & 0.5050 \\
        & Halstead Difficulty   &  0.3240 & 0.5310 \\
        & Halstead Effort       &  0.2499 & 0.6330 \\
        & Cyclomatic Complexity &  0.3469 & 0.5006 \\
        & NLOC                  & -0.1693 & 0.7485 \\
        \hline
    \end{tabular}
    \caption{Correlation coefficients between complexity metrics and F1 score.}
    \label{tab_correlation_coefficients} 
\end{table}

\subsubsection{\textbf{Visualization of Results}}

\begin{figure}
    \centering
    \includegraphics[width=1\linewidth]{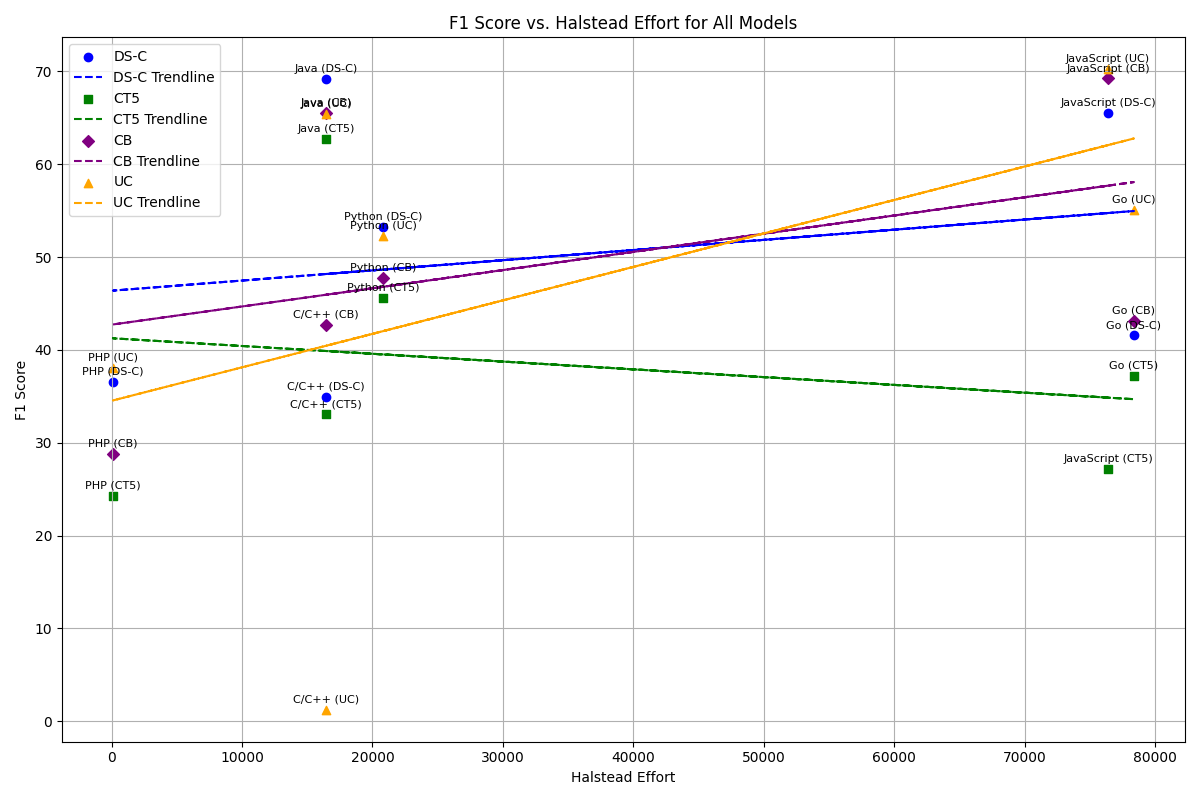}
    \caption{Mean Halstead effort vs. F1 score for each programming language.}
    \label{fig_halstead}
\end{figure}

\begin{figure}
    \centering
    \includegraphics[width=1\linewidth]{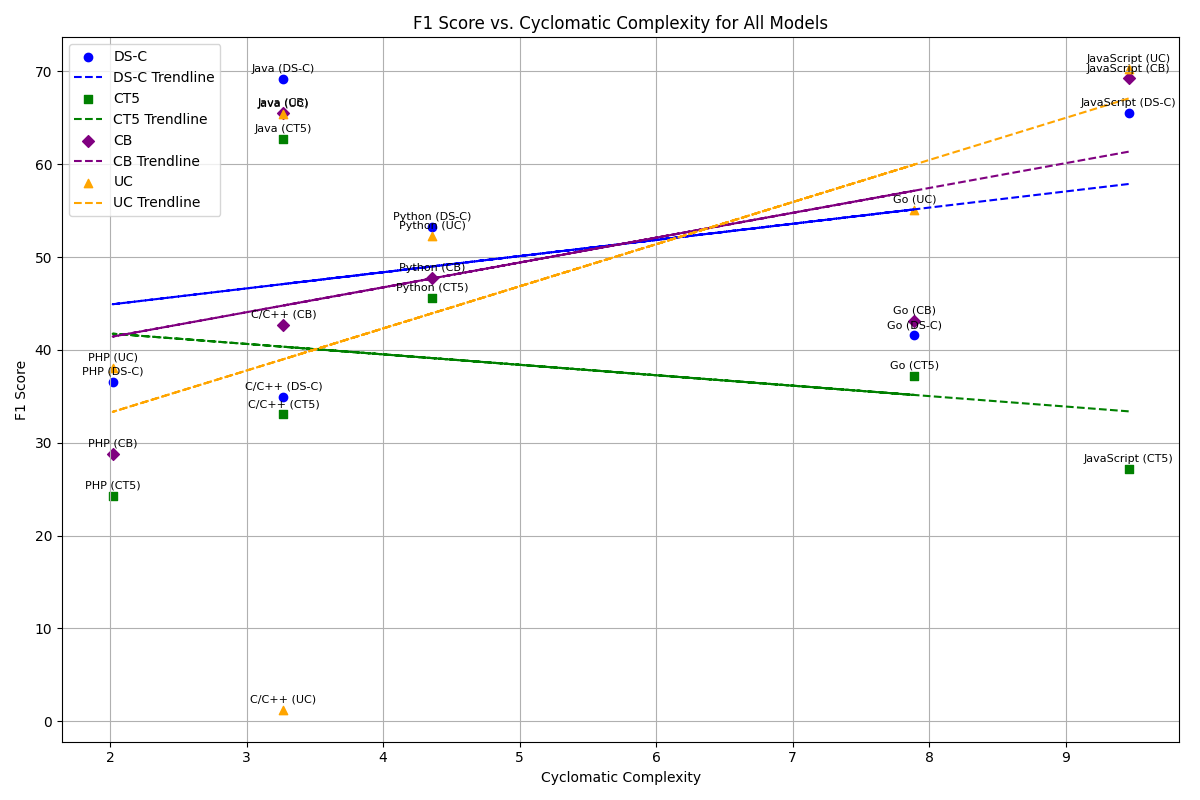}
    \caption{Mean cyclomatic complexity vs. F1 score for each programming language.}
    \label{fig_cyclomatic}
\end{figure}

To further explore the relationships, we plot the mean F1 scores against key complexity metrics, specifically Halstead Effort and Cyclomatic Complexity. In Figure~\ref{fig_halstead}, which shows F1 scores versus Halstead Effort, three of the trendlines (DS-C, CB, and UC) have a positive slope, while the CT5 trendline slopes downward. This mixture of upward and downward trends, along with the dispersed data points, suggests that no uniform relationship emerges. Similarly, Figure~\ref{fig_cyclomatic} plots F1 scores against Cyclomatic Complexity, and here again, DS-C, CB, and UC trendlines slope upward, whereas CT5 slopes downward. Overall, these mixed directions and the wide spread of points indicate that any correlation between these complexity metrics and F1 scores is weak and inconsistent across the different models.

\section{Limitation}
\label{threats}
Our study relies on the CVEfixes dataset, which is comprehensive and diverse, covering a significant portion ($\sim$80\%) of the available vulnerability fixes from the NVD database. While the CVEfixes dataset as a whole is extensive, the dataset sizes for individual languages like Go, Java, and Python are relatively small compared to the C/C++ datasets used in previous studies, such as DiverseVul and PrimeVul. These smaller dataset sizes are a result of the inherent distribution of vulnerabilities across languages in the NVD database (real-world representation) and not a limitation of the CVEfixes dataset itself. The CVEFixes paper acknowledges that at the time of writing,  $\sim$80\% of fixes in the NVD database refer to code on one of the three major forges: GitHub, GitLab, and Bitbucket, with 98\% of those being on GitHub. The remaining 20\% of fixes point to other forges (i.e., SourceForge or the defunct Gitorious) or to project-specific servers hosting other versioning systems (e.g., Mercurial, Subversion, or CVS). One could expand the dataset by manually locating and collecting data from projects that are not listed on GitHub, GitLab, or Bitbucket. This process would require manual collection and verification efforts.

However, even without the code hosted on non-GitHub, GitLab, and Bitbucket, we believe our investigation's results would not change significantly, as they only constitute the remaining $\sim$20\% of the total data.

\section{Conclusions}
\label{sec_conclusions}
In this study, we investigated the effectiveness of multiple language models (LMs) for vulnerability detection across several popular programming languages, including JavaScript, Java, Python, PHP, Go, and C/C++. Our results demonstrate that the performance of LMs varies depending on the language, with models generally performing better on non-C/C++ languages compared to previous work on C/C++. We found that vulnerability detection models for languages such as JavaScript and Java achieve higher F1 scores compared to the results reported in prior studies on C/C++. This suggests that LMs can be more effective and practical for vulnerability detection in these languages. However, performance varies between different languages, with PHP and Go showing high accuracy but lower F1 scores.

We also explored the relationship between code complexity and vulnerability detection performance across the studied languages. Our analysis revealed weak and statistically insignificant correlations between various complexity metrics and the models' F1 scores. This suggests that code complexity, as measured by these metrics, may not be a strong predictor of LMs' vulnerability detection performance.

We evaluated the generalizability of our findings on independent datasets for several languages. For Java, results on the MegaVul dataset closely mirrored those obtained from CVEFixes, suggesting consistent performance across different data sources. In contrast, testing C/C++ models on the PrimeVul dataset reaffirmed the difficulty of achieving robust vulnerability detection performance in these languages. For Python, performance on the synth-vuln-fixes dataset showed mixed results, indicating that some models struggle to generalize effectively. Similarly, testing the PHP models on the SARD dataset revealed challenges in detecting synthetic, well-characterized vulnerabilities as opposed to the more diverse real-world samples found in CVEFixes.

Our study has limitations that should be acknowledged. While the CVEfixes dataset is comprehensive, covering 80\% of the available vulnerability fixes from the NVD database, the dataset sizes for individual languages like Go, Java, and Python are relatively small compared to C/C++. Expanding the dataset to include the remaining 20\% of fixes from other platforms would require manual collection and verification efforts. However, we believe that the absence of this data does not significantly impact our findings.

\bibliographystyle{splncs04}
\bibliography{tex/biblio}
\end{document}